\documentclass{NAV}

\usepackage{boites}
\usepackage{amsmath,graphicx,color}
\usepackage{latexsym,amssymb,mathrsfs,amsfonts}	
\usepackage{framed,tikz}	
\usepackage{algorithm, algorithmic}
\usepackage{siunitx}
					
\usepackage[flushleft]{threeparttable}

\newcommand{\N}{\mathcal{N}}

\graphicspath{{figures/}}

\begin{document}

\title[Fusion of Classical and Quantum Accelerometers]{Enhancing Inertial Navigation Performance via Fusion of Classical and Quantum Accelerometers}

\author[]{Xuezhi Wang$^1$, Allison Kealy$^1$, Christopher Gilliam$^2$,  Simon Haine$^3$, John Close$^3$, Bill Moran$^4$, Kyle Talbot$^1$, Simon Williams$^4$, Kyle Hardman$^3$, Chris Freier$^3$, Paul Wigley$^3$, Angela White$^3$, Stuart Szigeti$^3$ and Sam Legge$^3$}

\address{\add{1}{School of Science, RMIT University, Australia}
\add{2}{School of Engineering, RMIT University, Australia}
\add{3}{Department of Quantum Science, Research School of Physics, Australia National University, Australia}
\add{4}{School of Engineering, University of Melbourne}
\email{xuezhi.wang@rmit.edu.au}}

\begin{abstract}
While quantum accelerometers sense with extremely low drift and low bias, their practical sensing capabilities face two limitations compared with classical accelerometers: a lower sample rate due to cold atom interrogation time, and a reduced dynamic range due to signal phase wrapping. In this paper, we propose a maximum likelihood probabilistic data fusion method, under which the actual phase of the quantum accelerometer can be unwrapped by fusing it with the output of a classical accelerometer on the platform. Consequently, the proposed method enables quantum accelerometers to be applied in practical inertial navigation scenarios with enhanced performance. The recovered measurement from the quantum accelerometer is also used to re-calibrate the classical accelerometer.  We demonstrate the enhanced error performance achieved by the proposed fusion method using a simulated 1D inertial navigation scenario. We conclude with a discussion on fusion error and potential solutions.
\end{abstract}


\maketitle

\section{Introduction}
\label{sec1}
Initially demonstrated in \cite{Carnal1991,Keith1991}, quantum accelerometers based on cold atom interferometry generate high precision measurements with extremely low drift over a long time period \citep{Bongs2019,Kitching2011,Degen2017}. Laboratory experiments demonstrate the accuracy of cold atom accelerometers to be fifty times greater than that of their classical counterparts \citep{Jekeli2005,Hardman2016}. This makes cold atom sensors potentially well-suited to inertial navigation systems, where the bias and drift of accelerometers and gyroscopes has a direct impact on the quality of positioning and attitude.

To deploy quantum sensors for inertial navigation, at least two challenges must be resolved \citep{2012:Aaron_Canciani}. First, the nature of cold atom interferometry is such that the resulting sample rate is typically below \SI{10}{Hz}. The lower the sample rate, the higher the achievable measurement accuracy \citep{Peters2001,Hardman2016}. This sample rate is significantly lower than existing classical accelerometers, which can operate at frequencies on the order of \SI{800}{Hz}.  Second, the dynamic range of quantum accelerometers is very low, typically reported below \SI{e-2}{\metre\per\second\squared} \citep{Gillot2014,Freier2016}. Fundamentally, the output signal of a quantum accelerometer is sinusoidal, with the body acceleration value proportional to the signal phase. When the body acceleration value is beyond the dynamic range, the output signal will be wrapped because of the circular phase of the sinusoidal wave.  In this paper we refer to this as the phase wrapping problem.

The phase wrapping problem was addressed by \cite{Bonnin2018}, who showed that implementing the simultaneous atom interferometers with different interrogation times could extend the dynamic range, whereas operating in phase quadrature improved the sensitivity. This approach increases the dynamic range of the quantum accelerometer at the cost of a more complicated hardware configuration. The work in \cite{Dutta2016} uses joint interrogation to eliminate the dead times, i.e. the preparation time between two adjacent cold atom interferometric cycles.

A hybrid cold atom/classical accelerometer was considered by \cite{Lautier2014}, where the atom interferometer signal phase is compensated by using the signal of a classical accelerometer. A similar idea is presented in \cite{Cheiney2018}, where a hybrid system combines the outputs of quantum and classical accelerometers in an optimization framework. An extended Kalman filter is used in the loop to estimate the bias and drift of a classical sensor as well as the phase of a quantum sensor. This results in a bandwidth of \SI{400}{Hz} and a stability of \SI{10}{ng} after $11$ hours of integration by the hybrid sensor. Although the accuracy of the quantum accelerometer was in the range of (\SIrange{0}{100}{\micro g}, simultaneously estimating the quantum sensor phase and classical sensor bias is nontrivial. Both the above two approaches are related to our work in the sense of  estimating the phase of a quantum accelerometer but differ in implementation and how the classical sensor bias is removed. More recent work by \cite{Tennstedt2020a} investigates the possibility of using the measurement of a cold atom interferometer sensor in Mach-Zehnder configuration in a navigation solution. They combine the cold atom sensor measurement with classical inertial sensors via a filter solution, observing an improved navigation performance when the underlying acceleration is small.

As mentioned above, existing efforts to extend the dynamic range of a quantum accelerometer essentially involves modifying the hardware configuration, which is nontrivial. In this paper, we propose a maximum likelihood probabilistic data fusion method that uses the accuracy of the quantum accelerometer --- operating at a low sample rate --- to re-calibrate a classical accelerometer over its full dynamic range. Our approach uses standard signal processing techniques and improves the inertial navigation capabilities of classical accelerometers without the need to extend the dynamic range of the quantum accelerometer. The idea is to unwrap the phase of the quantum accelerometer output signal by fusing the acceleration measurement of the classical sensor into the quantum sensor model at the quantum sensor sample rate.

The paper is arranged as follows. The problem statement and fusion idea are described in Section \ref{sec2}. The approach for quantum sensor signal phase unwrapping using a classical accelerometer via maximum likelihood estimation is presented in Section \ref{sec3}. The performance of the proposed method is demonstrated by simulation results in a 1D inertial navigation scenario in Section \ref{sec4}, which is followed by the conclusions in Section \ref{sec5}.

\section{Limitations of quantum accelerometers}
\label{sec2}

A quantum accelerometer operates by transforming a cloud of cold atoms into two spatially separated clouds in free fall such that the change of their vertical displacement in time mimics the two arms of an interferometer. The manipulation of the atom cloud is achieved by using one laser pulse to split the cloud and then a second to recombine the cloud. When the sensor has been subject to a specific force (e.g. acceleration and/or gravity), the two clouds of atoms exhibit different phase characteristics that can be measured. Counting the number of atoms in each cloud yields the relative phase, which in turn yields acceleration of the platform relative to the inertial frame defined by the freely falling atoms. Such a sensor has the potential to produce very precise measurements of acceleration or gravity \citep{Freier2016}. However, the sensor suffers from two key limitations that must be overcome.

The first challenge is that, in general, the quantum sensor has a low sampling rate that is governed by two features: the time required to produce a cloud of cold atoms, and the time that the atoms spend in free fall. The larger the size of the atom cloud and the longer these atoms spend in free fall, the better the precision of the acceleration measurement. A trade-off therefore exists between the performance of the sensor and the time between measurements. Some laboratories are exploring this trade-off to enable sample rates of up to \SI{330}{\hertz} \citep{butts2011light, mcguinness2012high}.

The second challenge the sensor faces is low dynamic range. Let $T$ be half the total interrogation time of the sensor, which is equivalent to half the time of flight for the atoms. For a constant acceleration $a$, the phase shift between the two atom clouds is \citep{Peters2001}
\begin{equation}\label{eq-1}
\triangle \phi = k_{\mathrm{eff}} a T^2,
\end{equation}
where the effective wave number is $k_{\mathrm{eff}} \approx 4\pi/\lambda$, and $\lambda$ is the wavelength of the laser. This phase shift is measured by the cold atom accelerometer as
\begin{equation}\label{eq-321}
S = N\sin(k_{\mathrm{eff}}aT^2 + \phi_0),
\end{equation}
where $N$ is the number of atoms in the atomic cloud and $\phi_0$ is the initial phase. Given $S$, an acceleration measurement is thus obtained by inverting \eqref{eq-321}. However, although a single acceleration $a$ uniquely determines $S$, the reverse is not true; a given $S$ can be obtained from distinct accelerations $a$.
\begin{figure}[htp!]
  \centering
  \includegraphics[width=0.6\textwidth]{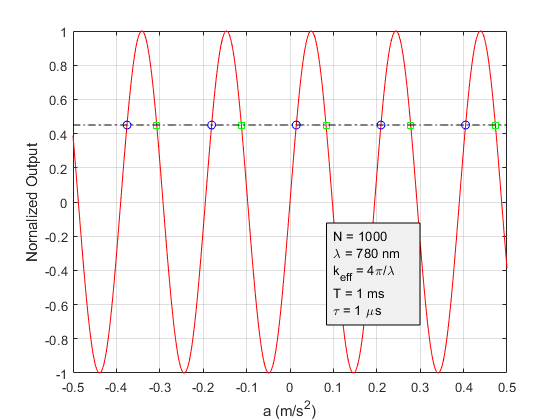}\\
  \caption{Normalized output signal ($S/N$) of the quantum accelerometer as a function of input $a$. An output signal (black dashed line) corresponds with multiple acceleration values as indicated by blue circles and green boxes, which may be the underlying acceleration measured.}\label{fig-01}
\end{figure}
Figure \ref{fig-01} illustrates the relationship \eqref{eq-321} with $N = 1000$ atoms, $\lambda = \SI{780}{\nano\metre}$, $T=\SI{1}{\milli\second}$, and a laser pulse width of $\tau = \SI{1}{\micro\second}$. We see that once the acceleration exceeds \SI{+- 0.05}{\metre\per\second}, an ambiguity occurs, and multiple accelerations map to the same $S/N$ value indicated by the black dashed line.

In this work, we are interested in unwrapping the phase of the accelerometer output $S$ to identify the underlying acceleration $a$. In Section \ref{sec3} we introduce an algorithm that performs this unwrapping by fusing a classical sensor with a quantum sensor. The resulting algorithm overcomes both the low sampling rate and low dynamic range challenges exhibited by the quantum sensor.

\section{Phase unwrapping by data fusion}
\label{sec3}

In this section we introduce our approach to unwrapping the phase of the quantum accelerometer via fusion with a classical device. Based on \eqref{eq-321}, we can write a complete noise-free quantum accelerometer  measurement model as follows \citep{Bonnin2018}:
\begin{align}
a &= f(S, s, n) \notag \\
&=\left\{\begin{array}{ll} \frac{s}{k_{\mathrm{eff}}T^2}\left[\arcsin(\frac{S}{N}) + 2n\pi + \phi_0 \right],& s = 1;\\ \frac{s}{k_{\mathrm{eff}}T^2}\left[\arcsin(\frac{S}{N}) - (2n+1)\pi-\phi_0 \right],& s = -1,  \end{array}\right. \label{eq-33}
\end{align}
where $s = \pm 1$ is the sign function and $n\in \mathbb{Z}$ is an integer.  In the presence of shot noise,  to a good approximation, this gives a signal model of the form \citep{2019:report_M1}
\begin{equation} \label{eq-34}
a_{\mathrm{out}} = f(S,s,n) + \nu_f,
\end{equation}
where $\nu_f \sim \N(0, \sigma_f^2)$, $\N(a,b)$ stands for a Gaussian distribution with mean $a$, variance $b$ and
$\sigma_f = 1/(k_{\mathrm{eff}}T^2\sqrt{N})$. Note that the two solution sets to \eqref{eq-33} are indicated by the green and blue circles in Figure \ref{fig-01}. Accordingly, to obtain the unwrapped acceleration we need to determine which equation to use and estimate the integer $n$.

We solve this dual estimation problem through fusion with a classical accelerometer. Our fusion algorithm is based on maximum likelihood estimation and comprises two steps. First, using the classical acceleration measurement $a_c$ and the output of the cold atom sensor $S$, a rough estimate of $n$ is obtained by inverting \eqref{eq-33} as follows:
\begin{subequations} \label{eq:roughn}
\begin{align}
    \hat{n}_1 &= \frac{k_{\mathrm{eff}}T^2}{2\pi}a_c - \frac{1}{2\pi}\left(\arcsin\left(\frac{S}{N}\right)-\phi_0\right), \\
    \hat{n}_2 &= \frac{k_{\mathrm{eff}}T^2}{2\pi}a_c + \frac{1}{2\pi}\left(\arcsin\left(\frac{S}{N}\right)-\phi_0\right) - \frac{1}{2}.
\end{align}
\end{subequations}
Note that $\hat{n}_1$ and $\hat{n}_2$ are rounded to the nearest integer. Second, the fused acceleration $a_f$ is obtained by evaluating \eqref{eq-33} for a finite set of integers centred around $\hat{n}_1$ and $\hat{n}_2$, and choosing the acceleration closest to $a_c$. The initial phase $\phi_0$ is a known constant determined by the hardware configuration. For simplicity, unless stated otherwise, we shall henceforth assume that $\phi_0 = 0$.

\begin{figure}[htp!]
  \centering
  \includegraphics[width=0.7\textwidth]{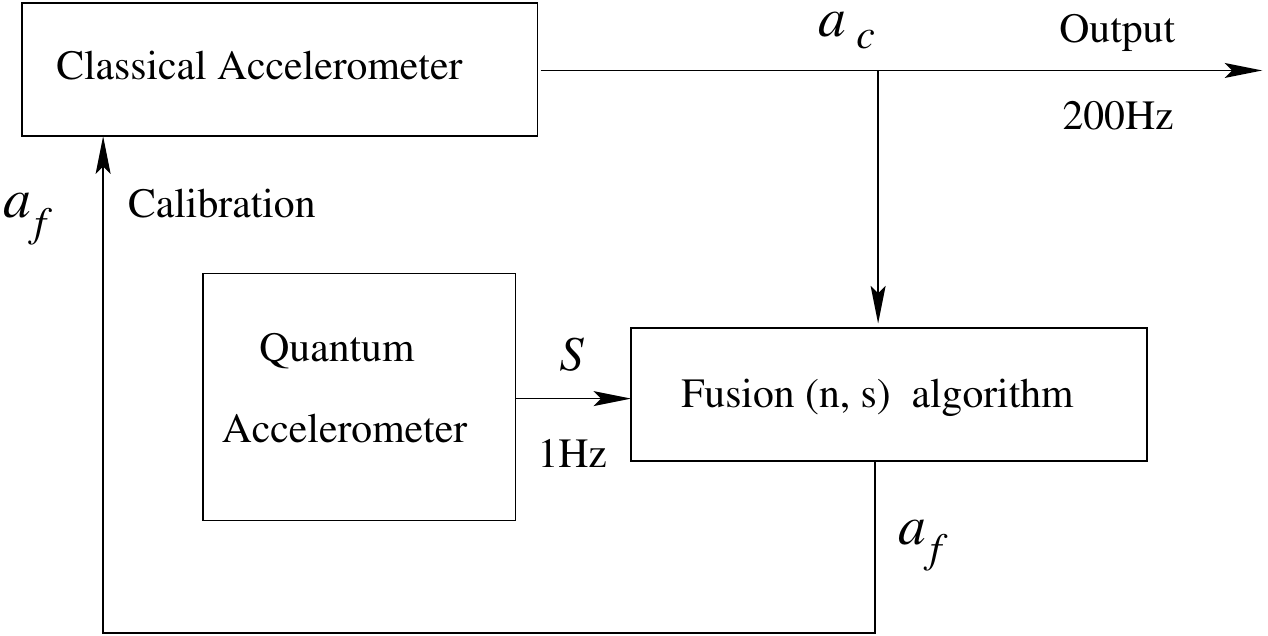}
  \caption{Fusion of classical accelerometer and quantum accelerometer for inertial navigation. The fused output $a_{f}$ re-calibrates the classical accelerometer output $a_c$ at \SI{1}{\hertz} (i.e., $a_c(k) \equiv a_{f}(k)$, where $k$ is the index of the quantum sensor sample time) and the latter is also the input of the fusion loop.}
   \label{f2}
\end{figure}

The fusion process, based on a maximum likelihood estimation approach, is specifically given by the following steps.
\begin{enumerate}
\item Input: $\hat{a}_c(k)$ from classical accelerometer; $S_k$ from quantum accelerometer.
\item Maximum likelihood method for parameter estimation:
\begin{equation}\label{mle}
(n_k^o, s_k^o) = \arg\max_{n_k \in \mathbb{Z}, \, s_k \pm 1} p\Bigl(a_{out}(k)|S_k,\hat{a}_c(k)\Bigr),
\end{equation}
where, following \eqref{eq:roughn},
\begin{align}
n^o_k &= \left\{\begin{array}{ll} \frac{k_{\mathrm{eff}}T^2}{2\pi}\hat{a}_c(k) -\frac{1}{2\pi}\arcsin(\frac{S_k}{N}), & s^o_k = 1; \\ \mbox{} \frac{k_{\mathrm{eff}}T^2}{2\pi}\hat{a}_c(k) + \frac{1}{2\pi}\arcsin(\frac{S_k}{N}) - \frac{1}{2}, & s_k^o=-1. \end{array}\right. \notag
\end{align}

\item Estimate $a_{f}(k)$, the fused output as shown in Figure \ref{f2}, given $S_k$ and $n_k^o$ and $s_k^o$:
\begin{align}
p\bigl(a_{\mathrm{out}}(k)|S_k,n^o_k,s^o_k\bigr) & = \N\bigl(a_{f}(k), \sigma_f^2\bigr), \notag
\end{align}
where $\sigma_f = 1/(k_{\mathrm{eff}}T^2\sqrt{N}).$

\item Convergence check:
$\hspace{0.5 cm} ||\hat{a}_{out}(k)-\hat{a}_{f}(k)|| \leq \varepsilon$. The classical accelerometer reading is then reset by the fused output $a_f(k)$ at $k$.
\end{enumerate}
Figure \ref{f2} illustrates this procedure.

We now evaluate the performance of the proposed algorithm via Monte-Carlo simulations under the condition that the dynamic range of ground truth acceleration is well beyond that of the quantum accelerometer but within that of the classical accelerometer. At each run, the data of ground truth acceleration is drawn from a uniform distribution. The following configuration for the cold atom sensor is used: $N = 1000$ atoms, $T = \SI{1}{\milli\second}$ half interrogation time, $\tau = \SI{1}{\micro\second}$ beam splitter pulse width of the laser, and a laser wavelength of $\lambda = \SI{780}{\nano\metre}$.

We assume that the output of the classical sensor can be expressed as
\begin{align}
a_{c}(t) &= a(t) + w(t), \label{eq51}
\end{align}
where $a(t)$ is the true acceleration and $w(t)$ represents the error associated with the sensor. In the literature, the latter is approximated by a sum of three independent Gaussian random variables \citep{Quinchia2013}:
\begin{equation} \label{eq:accelnoise}
w(t) \sim \N(b, \sigma_q^2+\sigma^2_{bo}+\sigma^2_{bd}\sqrt{t}).
\end{equation}
Here $\sigma_q$ is the standard deviation of precision error, modeled as white noise; $\sigma_{bo}$ denotes the standard deviation of bias offset, modeled as a first order Gauss-Markov process; and $\sigma_{bd}$ denotes the standard deviation of bias drift, modeled by a Gaussian random walk. $b$ signifies constant bias and its value is set to \SI{2e-3}{\metre\per\second\squared} in the simulation, equivalent to the bias of a tactical grade sensor \citep{Chow2011}.

\begin{figure}[htp!]
\centering
\includegraphics[width=0.6\textwidth]{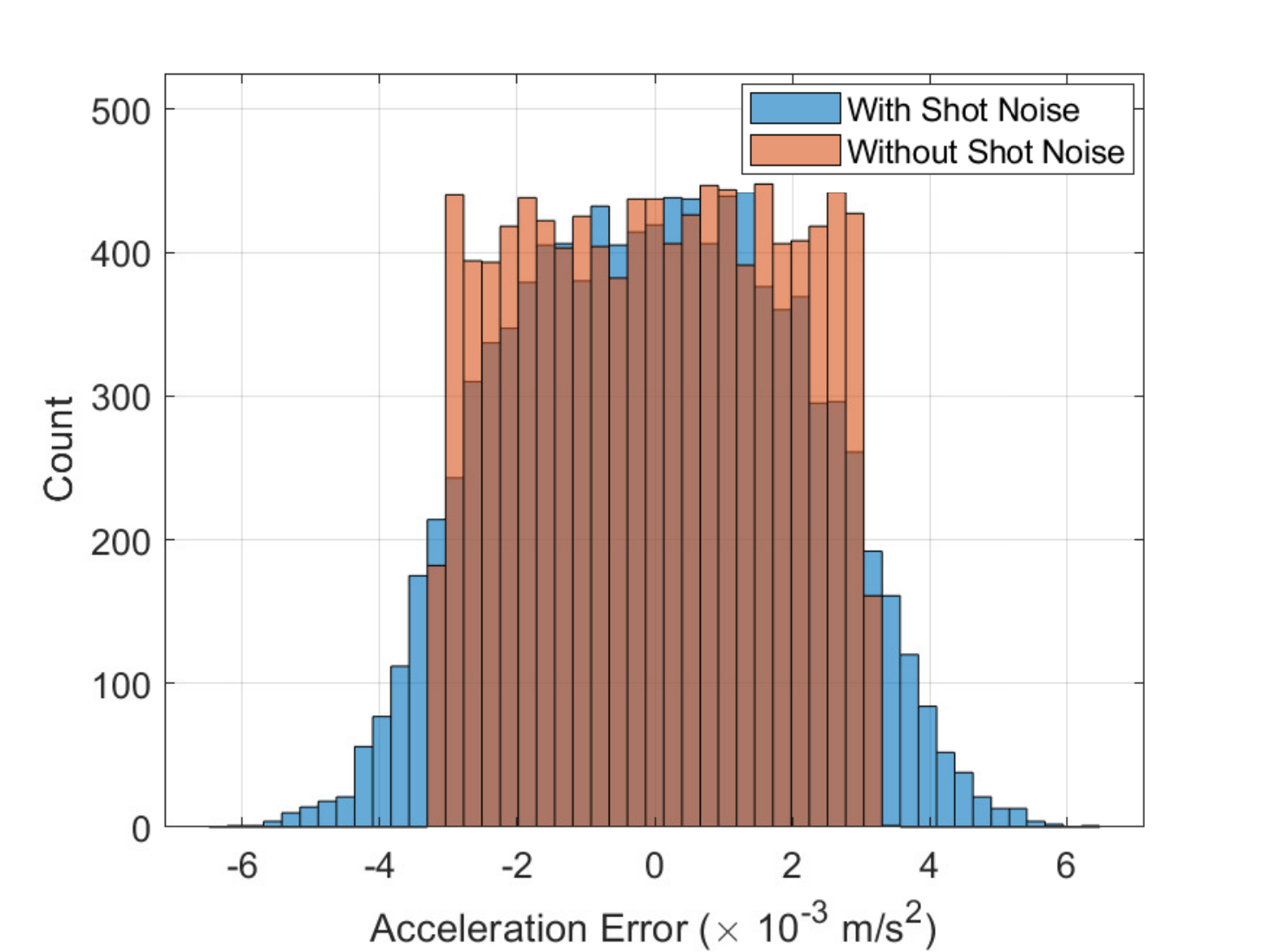}
\caption{Fusion error statistics from 10,000 Monte-Carlo runs for
$a$ drawn from $\mathcal{U}(\SI{-10}{ms^{-2}},\SI{10}{ms^{-2}})$ in the presence (blue) and absence (orange) of shot noise.}
\label{fig-05}
\end{figure}

\begin{figure}[htp!]
\centering
\includegraphics[width=0.6\textwidth]{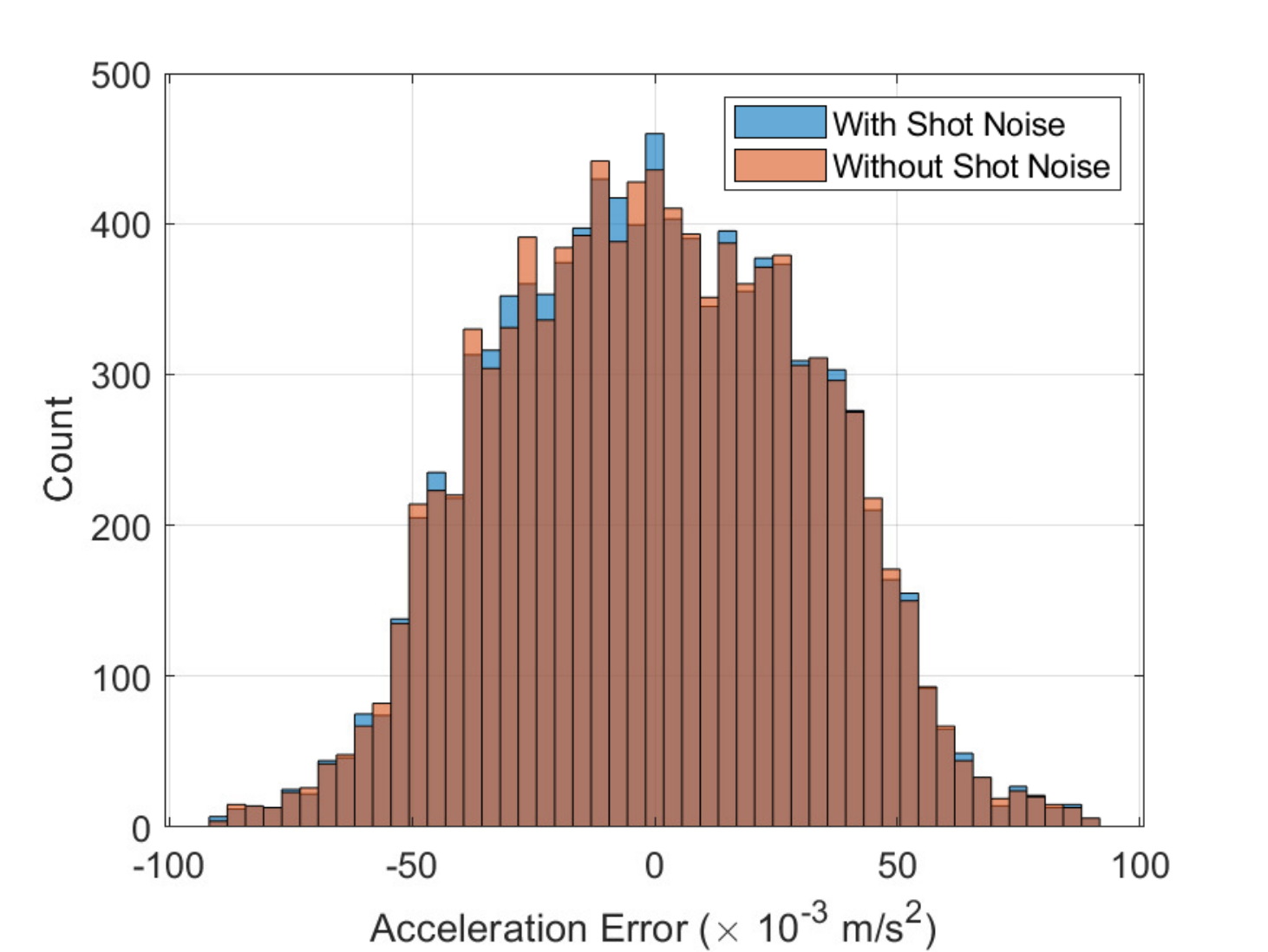}
\caption{Fusion error statistics via 10,000 Monte-Carlo runs for $a$ drawn from $\mathcal{U}(\SI{-e3}{ms^{-2}},\SI{e3}{ms^{-2}})$ in the presence (blue) and absence (orange) of shot noise.}
\label{fig-06}
\end{figure}

The error statistics for this evaluation are shown in Figure \ref{fig-05} and Figure \ref{fig-06}, based on 10,000 Monte-Carlo simulations. These histograms show the error between the fused acceleration estimate and the original classical measurement for two situations: the cold atom sensor with shot noise corruption (blue) and without (orange). In Figure \ref{fig-05}, the classical acceleration is drawn from $\mathcal{U}(\SI{-10}{ms^{-2}},\SI{10}{ms^{-2}})$ whereas in Figure \ref{fig-06} the measurement is drawn from $\mathcal{U}(\SI{-e3}{ms^{-2}},\SI{e3}{ms^{-2}})$. We observe that the main contribution to the fusion error spread without shot noise (orange) is the nonlinear sensitivity of the mapping between the sensor output signal and the underlying acceleration.  This error is discussed further in Section \ref{sec4}.

In the next section, we demonstrate the performance of our fusion approach using simulations of a one-dimensional inertial navigation scenario.

\section{Navigation simulation results}
\label{sec4}

\begin{figure}[htp!]
  \centering
  \includegraphics[width=0.6\textwidth]{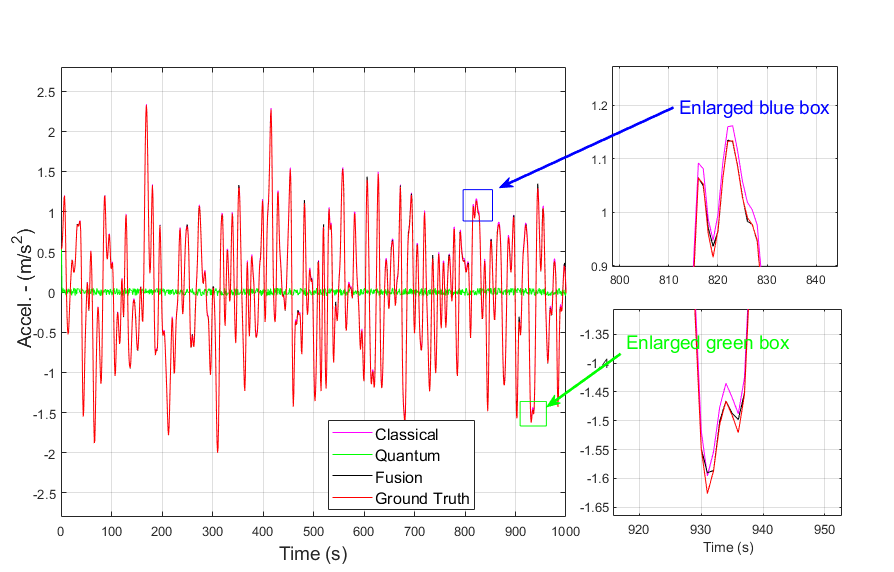}\\
  \caption{Comparison of acceleration values between classical, quantum, fusion and ground truth in a single run. Apart from the quantum accelerometer output which indicates in green, the other three are overlapped in this figure. We highlight the error difference between the classical and fusion with respect to ground truth in Figure \ref{fig-071}. }\label{fig-07}
\end{figure}

Consider a vehicle moving along a single axis with an onboard strapdown inertial navigation system. The system consists of a classical accelerometer and a quantum accelerometer. The initial position and velocity of the vehicle are set to zero. As shown in Figure \ref{fig-07}, the underlying body acceleration randomly varies with time, and its spectrum follows a zero-mean Gaussian distribution $a \sim \N(0, \sigma_a^2)$ with standard deviation $\sigma_a = \SI{1}{\metre\per\second\squared}$.

As in \eqref{eq:accelnoise}, the measurement noise of the classical accelerometer is given by a sum of three Gaussian distributions: precision error $\sim \N(0, \sigma_b^2)$; bias offset $\sim \N(b_1,\sigma_{b_1}^2)$; and bias drift $\sim \N(b_2,\sigma_{b_2}^2 \sqrt{t})$, where bias drift is modeled as a Gaussian random walk and we chose $b_1 = b_2 = \SI{e-3}{\metre\per\second\squared}$ and $\sigma_{b_1} = \sigma_{b_2} = 1$. The measurement noise of the quantum accelerometer is dominated by shot noise with distribution $v_f \sim \N(0, \sigma_s^2)$, where $\sigma_s = 1/(k_{\mathrm{eff}}T_{pi}^2\sqrt{N})$.
The effective wave number of the quantum accelerometer is $k_{\mathrm{eff}} = 4\pi/\lambda$, $\lambda = \SI{780}{\nano\metre}$, the half interrogation time $T = \SI{1}{\milli\second}$, the duration of the laser beam splitter is assumed to be $\tau = \SI{e-6}{\second}$ and the average number of atoms per shot is $N=1000$. The sampling rate of the classical accelerometer is \SI{200}{\hertz}, while the quantum accelerometer is set at \SI{1}{\hertz}.

In this simple strapdown navigation scenario we examine two cases. In the first, navigation is driven by the output of the classical accelerometer alone, and in the second, navigation is driven by the fusion of the classical with the quantum, where the classical accelerometer is calibrated at the sampling rate of the quantum accelerometer by the output of the fusion procedure. We compare the navigation errors in the two cases.

\begin{figure}[htp!]
  \centering
  \includegraphics[width=0.6\textwidth]{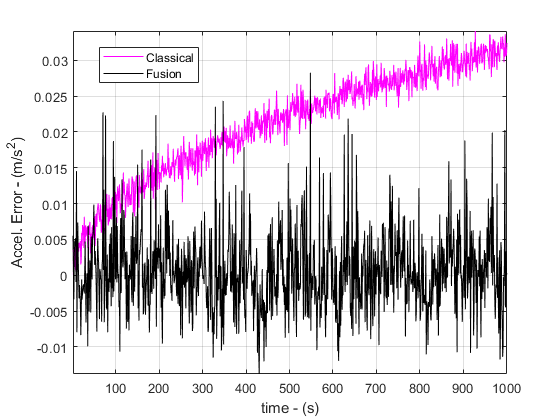}\\
  \caption{Illustration of the error difference of output values between classical and fusion accelerometers with respect to ground truth.}\label{fig-071}
\end{figure}

\begin{figure}[htp!]
  \centering
  \includegraphics[width=0.6\textwidth]{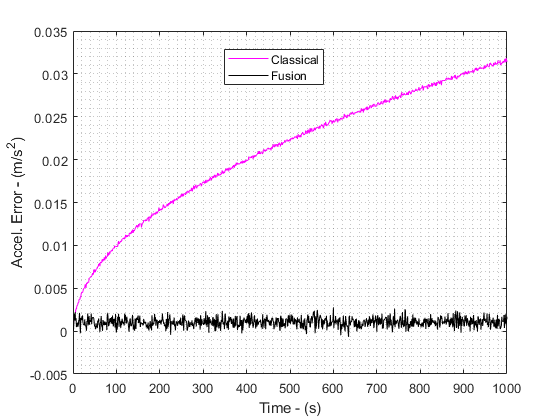}
  \caption{Acceleration errors vs time averaged over 1000 runs. }\label{fig-09}
\end{figure}

\begin{figure}[htp!]
  \centering
  \includegraphics[width=0.6\textwidth]{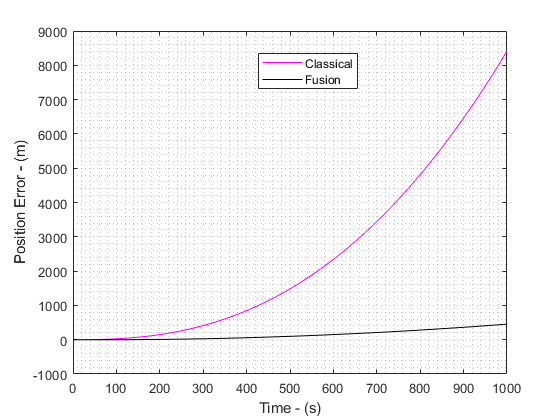}
  \caption{Position errors vs time averaged over 1000 runs. }\label{fig-10}
\end{figure}

Figure \ref{fig-07} shows the comparison of estimated and ground truth acceleration values over $1000$ seconds, where the curves inside the blue and green boxes are enlarged to highlight their differences. The error difference between classical and fusion measurements with resepct to ground truth are illustrated in Figure \ref{fig-071}.
It can be seen that the measurement of the quantum accelerometer cannot follow the ground truth acceleration caused by the signal with wrapped phase (Figure \ref{fig-07}), and the fused acceleration measurements yield a smaller error than that of the classical accelerometer alone (Figure \ref{fig-071}).  To demonstrate that our proposed fusion process extends the dynamic range of the quantum accelerometer, we compute the averaged acceleration error and position error from 1000 Monte-Carlo runs driven by a random acceleration profile, as shown in Figures \ref{fig-09} and \ref{fig-10}.  Figure \ref{fig-09} shows that the acceleration reading error of the classical accelerometer is accumulated over the duration, while the fusion algorithm is able to remove the accumulated bias by regularly re-calibrating the reading of the classical accelerometer at the sampling rate of the quantum accelerometer. As shown in Figure \ref{fig-10}, the fusion result demonstrates a substantial improvement in position error performance through the data fusion process over using the classical accelerometer alone.

Our experiments show that the fused acceleration error exists, and it increases with the scale of the underlying body acceleration.  One major contribution to the fusion errors is from the mapping between the output signal $S$ and input acceleration $a_q$ of the quantum accelerometer, which presents different sensitivities due to the nonlinear nature of the sine function. Reading $a_q$ from the linear part of the sine function in $S$ can address the problem. This was shown by \cite{Bonnin2018}, who used two orthogonal phased quantum accelerometers to remove the nonlinear sensitivity problem of the quantum accelerometer. 

\begin{figure}[htp!]
  \centering
  \includegraphics[width=0.6\textwidth]{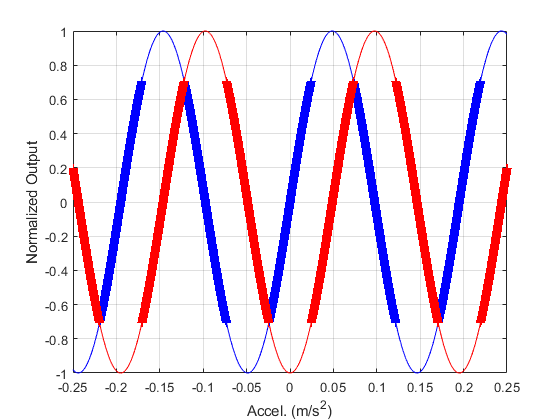}
  \caption{Illustration of the normalised output signals of two orthogonal phased quantum accelerometers vs acceleration. The bold curve highlights the parts of linear sensitivity from the two sensors across the acceleration range.  $\phi_1 = 0$ (red) and $\phi_2 = \pi/2$ (blue). }\label{fig-11}
\end{figure}

As shown in Figure \ref{fig-11} --- a normalised plot for the expected output signal versus the value of acceleration --- we assume that two ``exactly orthogonal phased'' quantum accelerometers are available for the fusion process.
At this stage, we assume that the output switching between the two quantum sensors is determined by selecting the normalised signal output which satisfies $S < N\sqrt{2}/2$, and that there is no switching error in the simulation.
\begin{figure}[htp!]
  \centering
  \includegraphics[width=0.75\textwidth]{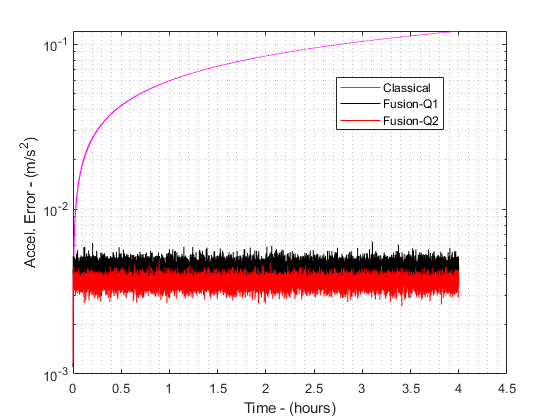}
  \caption{Comparison of Root-Mean-Squared acceleration error averaged over 1000 runs in the 4-hour scenario. `Classical': output of classical accelerometer; `Fusion-Q1': output of classical and quantum accelerometer fusion; `Fusion-Q2': same as `Fusion-Q1' but using orthogonal phased quantum accelerometers. }\label{fig-12}
\end{figure}
\begin{figure}[htp!]
  \centering
  \includegraphics[width=0.75\textwidth]{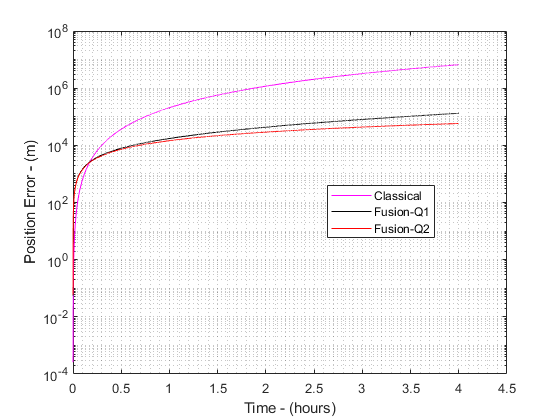}
  \caption{Comparison of Root-Mean-Squared position error averaged over 1000 runs in the 4-hour scenario.}\label{fig-13}
\end{figure}

We repeated the Monte-Carlo simulation shown in Figures \ref{fig-09} and \ref{fig-10} with a navigation period of $4$ hours, with the new simulation results shown in Figures \ref{fig-12} and \ref{fig-13}, respectively. These plots demonstrate that by avoiding the use of the nonlinear part of the quantum accelerometer for fusion output, there is almost no drift remaining in the fused acceleration error (see the curve `Fusion-Q2'), which is dominated by shot noise.

As ongoing research, we will continue our investigation along these lines for handling phase noise between the two orthogonal-phased quantum accelerometers, and simulating an adaptive phase-locked loop implementation for inertial navigation systems of predictable accelerations.

\section{Conclusions}
\label{sec5}

This paper proposes a fusion method that extends the dynamic range of a quantum accelerometer by unwrapping the signal phase of the quantum accelerometer output from the reading of a classical accelerometer using a maximum likelihood estimator. Consequently, the fusion process removes accumulative drift of the classical accelerometer by re-calibration using fusion output and thus enables the classical sensor to gain a substantially reduced drift over a dead reckoning navigation process. Promising performance is observed in the simulation results presented.

\bibliographystyle{unsrtnat}
\bibliography{ICASSP2020}

\end{document}